\documentclass[12pt,a4paper]{cibb}
\usepackage{fancyhdr}
\usepackage{subfigure,graphicx}
\usepackage{amsmath,amsfonts,latexsym,amssymb,euscript,xr}
\usepackage{booktabs}
\usepackage[nodayofweek]{datetime}
\usepackage{hyperref}
\usepackage[table]{xcolor}
\usepackage{color,colortbl,tabularx}
\usepackage[english]{babel}
\usepackage[protrusion=true,expansion=true]{microtype}
\usepackage{amsmath,amsfonts,amsthm}
\usepackage{pifont}
\usepackage[numbers]{natbib}
\usepackage{comment}

\graphicspath{{../pdf/}{../jpeg/}}
\DeclareGraphicsExtensions{.pdf,.jpeg,.png}

\usepackage{mathabx}
\usepackage{algorithmic}
\usepackage{array}
\usepackage{mdwmath}
\usepackage{mdwtab}
\usepackage{eqparbox}
\usepackage{dirtytalk}
\usepackage{csquotes}
    \usepackage{float}
    \usepackage{csquotes}

\usepackage{url}
\definecolor{LightBlue}{rgb}{0.88,0.9,0.9}

\title{\Large $\ $\\ \bf  Towards a Personal Health Knowledge Graph Framework for Patient Monitoring}

\author{\large Daniel Bloor$^{1}$, Nnamdi Ugwuoke$^{1}$, David Taylor$^3$, Keir Lewis$^{4}$, Luis Mur$^{2}$, Chuan Lu$^{1}$}
\address{\footnotesize $\ $\\
$^1$ Department of Computer Science, Aberystwyth University, Aberystwyth, UK. \{dab45, nnu1, cul\}@aber.ac.uk\\
$^2$ Department of Life Science, Aberystwyth University, Aberystwyth, United Kingdom.\\ 

$^3$ Valley Diagnostics Limited, Cardiff, United Kingdom. \\
$^4$ Biomedical Sciences, Swansea University, Swansea, United Kingdom. \\

}

\abstract{\small \textit{personal health knowledge graph, patient monitoring, COPD} \normalsize
\\[17pt]
{\bf Abstract.} Healthcare providers face significant challenges with managing and monitoring patient data outside of clinics, particularly with limited resources and insufficient feedback on their patients' conditions. Effective management of these symptoms and exploration of larger bodies of data are vital for maintaining long-term quality of life and preventing late interventions. 
In this paper, we propose a framework for constructing personal health knowledge graphs from heterogeneous data sources. Our approach integrates clinical databases, relevant ontologies, and standard healthcare guidelines to support alert generation, clinicians' interpretation and querying of patient data. Through a use case focusing on monitoring Chronic Obstructive Lung Disease (COPD) patients, we demonstrate that inference and reasoning on personal health knowledge graphs built with our framework can aid in patient monitoring and enhance the efficacy and accuracy of patient data queries.}

\begin{document}

\thispagestyle{fancy}
\pagestyle{fancy}
\fancyhead{} 
\renewcommand{\headrulewidth}{0pt}
\fancyhead[L]{\small \texttt{Proceedings of the 18th Conference on Computational Intelligence\\Methods for Bioinformatics \& Biostatistics (CIBB 2023)}}
\fancyfoot{} 
\fancyfoot[C]{\thepage}

\section{\bf Introduction}
\label{sec:SCIENTIFIC-BACKGROUND}

Personal health knowledge graphs (PHKG) are sophisticated data representations of medical concepts and the relationships between them, allowing researchers to integrate and analyze patient data from various sources, including electronic health records (EHRs), claims, and unstructured clinical notes \cite{phkg} \cite{PHKG-Diet}. PHKGs are powerful approaches to linking domain knowledge and the relevant terminology with clinical data for patients. They can be used to create subgraphs in patient querying, which can aid in clear and specific visualizations and efficiency. PHKGs also provide the necessary connections and relationships between terminology to provide appropriate context within the domain model. This, in turn, augments the patient data for a broader understanding and interpretation.

A conceptual solution was proposed in \cite{phkg} for combining IoT data analytics and explicit knowledge upon 3 chronic disease use cases, where PHKGs are managed using an RDF triple store. Other related works have explored the use of ontology-based data access solutions. For example in OntoMongo \cite{ontomongo}, an RDF ontology mapping layer is used to connect the data stored in MongoDB, a popular NoSQL document store, enabling SPARQL query over data through query translation. 
PrimeKG was created as a multimodal knowledge graph for precision medicine, integrating diverse resources to support artificial intelligence driven research in personalized diagnostics and treatments \cite{PrimeKG}.

Despite the progress made, research in the development of PHKGs is still in its infancy. Here, we propose a framework of constructing PHKGs for chronic disease monitoring, a task that presents challenges for both clinicians and patients, particularly outside of clinical settings. Our framework aims to integrate expert knowledge and multimodal personal health and environmental data, such as sensory data, EHRs, omics and imagery data, to generate actionable insights and alerts. Various data transformation and harmonization steps are proposed. Semantic technologies, including natural language processing (NLP), ontologies, and machine learning, will be used for annotation and mapping for creating nodes and relationships in the knowledge graphs.

Based on this framework, we implemented a modular system as a use case for monitoring patients with chronic obstructive pulmonary disease (COPD). It relies on medical ontologies, healthcare guidelines and the use of MongoDB and Neo4j, and a reasoning engine enabling rule-based and model-based inferences. Raw data and all sorts of metadata are stored using MongoDB. Whilst the general KG and the PHKGs are managed using Neo4j, which is a type of property graph database that has benefits over RDF and SPARQL in terms of schema flexibility and query expressiveness. Neo4j also has an easy-to-understand Cypher query language and allows for more information to be defined within the nodes and relationships in the graph itself. Our prototype system represents an initial step in the development PHKGs for COPD. It is adaptable in the future by enabling more machine learning capability. This can help fill the gaps in the PHKGs and enhance the reasoning capability for the system.

\section{\bf Methods}
\label{sec:DATA-AND-METHODS}

\subsection{Data}
The MIMIC-III\cite{mimiciii} dataset was used to test the prototype system and validating some components of the proposed framework. MIMIC-III\cite{mimiciii} is a large freely accessible database featuring anonymized data on patients admitted to critical care units at a tertiary care hospital. This gave us a multimodal dataset of both structured and unstructured data, including vital sign measurements and clinical notes, for a large number of patients with varying symptom severity. We used this data to test our risk scoring and alert generation for COPD patients, and it demonstrated the challenges of managing such data. Although our system is designed to be as independent of specific data sources as possible.

\subsection{Proposed framework and implementation}

\begin{figure}[ht]
    \centering
    \includegraphics[width=1\textwidth]{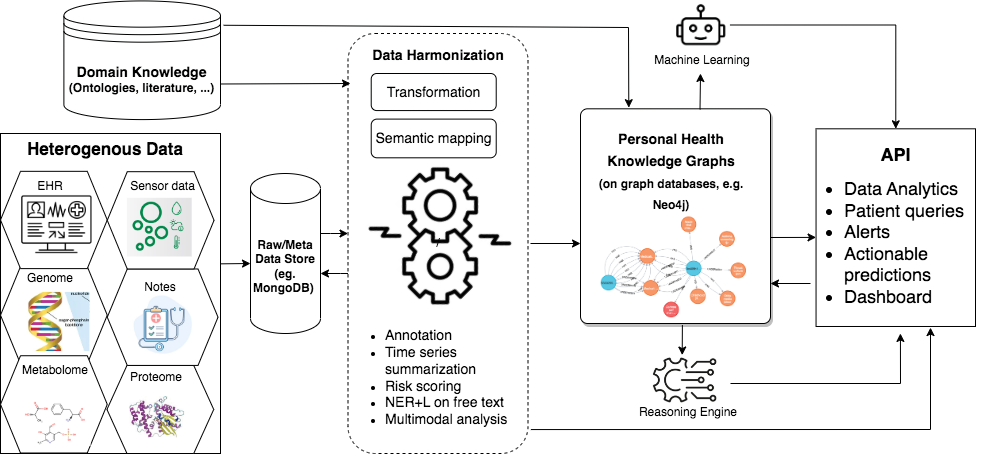}
    \caption{The personal health knowledge graph (PHKG) framework}
    \label{fig:architecture}
\end{figure}

The proposed framework consists of the following main components (see Figure \ref{fig:architecture} for the overall architecture): 1) Selection, preparation and merging of ontologies for relevant disease management; 2) Data processing for transformation, annotation of different data streams as well as the associated metadata; 3) Constructing general knowledge graph (KG) from the ontology, adding rules/ risk scoring systems based on the literature and inferences from the ontology's expert knowledge; 4) Creating nodes (vertices) and relationships (edges) from the transformed data by mapping them to the ontological concepts and storing them in a graph database, e.g. Neo4j; 5) Reasoning/Inferences on the graph data using rules based reasoning or model based inferences;  6) Extract personalised health KGs by retrieving relevant subgraphs, update or augment the PHKGs with new fact or prediction from the reasoning/inference; 7) API for downstream tasks on the PHKGs. 

\textbf{1) Ontologies. }
 For the use case, an ontology was developed to store the expert knowledge semantic terminology relevant to monitoring COPD symptoms and patient condition. The ontology constructed for the project is known as COPH (COpd and Physical Health). This is formed of multiple ontologies, imported and merged in Protege\cite{protege}, that have different discrete domain knowledge (ECG, ICD9 codes, diseases etc.) that are collectively relevant to perceived use cases for the platform as whole. This ontology contains 106,242 classes considered relevant for managing COPD and the associated physical health context. Snomed-CT ontology is also included for annotation of free text. 

\textbf{2) Data harmonization.} \\
\textit{Time series data summarization.} This is an important step for constructing temporal KGs. For example, in our use case of MIMIC-III data, the summaries (such as averages and the mode) of the measurements are splitting a day into quarters, i.e., a summary for each 6 hours of the day. Timestamps are stored typically as properties of a node and can be accessed easily through queries, and smart filters can be applied using comparisons with other dates, such as only returning results featuring a timestamp within a time/date-range.

Statistics can also be computed for detecting condition shifts, e.g. the Kolmogorov-Smirnov test can be used to compare the distribution of the most recent 3 days of the patient's measurements with the rest of the month before it in our use case. This comparison can help to understand whether something concerning may be happening with the patient's symptoms acutely.

\textit{Dealing with free text.} Appropriate language models can be utilized for extract and annotate the free text into key clinical concepts and relationships, often model fine-tuning might be needed for the target domain. Here we utilized the Medical Concept Annotation Toolkit (MedCAT), which is an open-source library providing Named Entity Recognition and Linking (NER+L) functionalities, as well as an annotation tool and an online learning training interface \cite{MedCAT}. 
For our use case, we initially extracted information from MIMIC III patient note in order to configure the vocab and concept database (CDB) for the MedCat model.

The vocab model has functions of spell checking and word embedding, which can be pre-calculated using word2vec or BERT. The concept database is a custom built using the chosen ontologies such as Snomed-CT. We then performed self-supervized training on the raw document to train or refine the MedCat models, which will be used for NER+L on the documents. The resulting annotation output would be imported into a graph database and merge with the PHKGs for corresponding patients. 

\textit{Harmonizing multimodal data.} Biomedical research and precision medicine increasingly rely on omics data, such as genomics and metabolomics, as well as medical images and documents. These diverse data types can be harmonized based on semantic summarization of clinical reports or predictions from multimodal biomedical AI models. Multimodal analysis can generate actionable insights, such as disease subtypes, biomarker profiling, and associated pathways \cite{mmAI2022}. These results can be annotated to the relevant concepts and then stored and queried conveniently using PHKGs. Relevant pretrained machine learning models can also generate embeddings for data items of different modalities. These embeddings can be stored in the PHKG, ready for graph indexing or inferences based on semantic similarity between nodes or subgraphs.

\textbf{3) General KG.}
General KG can be constructed directly from ontologies: we create the instances for the classes from those Internationalized Resource Identifiers (IRIs) featured in the ontology and create relationships between the measurements and the patient (who have their own instances in the knowledge graph). These classes, known as nodes for the knowledge graph, and relationships can then both be written to the knowledge graph.

Additionally, scoring systems for determining risk from measurements or rule-based models for risk assessment can be implemented based on standard healthcare guidelines. These models are explainable, which can help to both trust the insights and understand where refinement and personalization can be added. The KG can be updated based on domain knowledge extracted from ontologies and the literature (including existing knowledge bases) \cite{PrimeKG}. 

\textbf{4) Personal health KG. }
To integrate the transformed data with the knowledge graph, a semantic mapping layer was implemented to annotate the data using metadata referencing ontological terms. Each class will have an appropriate object property with a theme of possession. For example, a class for ``Condition'' (diagnosis) would have a corresponding object property ``hasCondition'' that will be used for the mapping. Appropriate node properties (such as a patient node having an age) would be a property of the node rather than another relationship. This is to keep a clean and efficient architecture that is easier to traverse. 

We use MongoDB to store the mapping with version control to support traceability and provide transparency for the nature of the terms and reduce definition confusion. The semantic mapping is part of our transformation layer and the resulting mapped data is ready for storage in the knowledge graph.

\textbf{5) Reasoning and Inference. }
A reasoning engine was implemented that can utilize both rule-based and model-based inference. Once we have the terms created for the data and the relationships in place, we can consider the terms to see what can be inferred from the data. Relationships such as subclass or ISA are commonly used in queries without explicitly listing associated subclasses in the query statement. 

\textbf{6) Personalization. } In the system for COPD, risk scores and higher level vital sign context were inferable through rules set into the graph database based on a combination of the ontology's expert knowledge and personalized thresholds. Cypher query triggers were set up to trigger upon committing the data to the knowledge graph. The knowledge graph can store both generalized and personalized thresholds and flags for comparison, which can then be updated for refining the personalized care for the patient. A model-based approach can also be utilized for inferring relationships, context or determining patient risk. These scores are evaluated to create alerts of varying grades to indicate severity and levels of urgency to the healthcare provider and patient for either seeking consultation or the provision of appropriate care.

\textbf{7) Data Storage. }
A major element of the data management system for the COPD use case, is being able to store the data that will be mapped to the ontology for further processing. MongoDB was chosen for its schemaless, flexible document-store which allows a more relaxed management approach, where both the raw multimodal data, all relevant metadata for data transformation and mapping would be stored. 

While this project is using MongoDB specifically for the raw data storage, the architecture as a whole can easily support an alternative database system and is designed to be agnostic to the specific tools, providing that the functionality provides the general inputs and outputs required for the previous and next components in the framework.

Neo4j's maturity compared to the alternatives made it an easy choice for working into the system for storing PHKGs without being concerned about potential complications, and it has plugins available for the features that may be missing. 
With timestamp attached to the PHKGs, and a snapshot or a subgraph of the PHKG at a specific timestamp can be easily extracted from the graph database. For time-series data streams, only summary or compressed data would be stored as graphs in order to keep the graph database leaner for faster queries, while also retaining the appropriate context of the patient's condition. 

A separate storage for the rest of the data, such as the MongoDB solution, makes more sense for the simple document structure that will be more memory efficient on larger datasets. Therefore there is a complementary relationship between the database solutions for supporting the framework overall.

\section{\bf Results}
\label{sec:RESULTS}

For the use case of COPD monitoring, we constructed a PHKG features 3.5 million nodes and 4 million relationships. We utilized the APOC library for Neo4j to create the triggers that will trigger upon committing new data to the graph database, where it will then execute the queries that assess for thresholds and ranges of measurement values deemed appropriate for the patients. Personalized thresholds are prioritized and if there is one available for the patient for the given measurement type, it will trigger in place of the general rule based on standard expert knowledge. 

\begin{figure}[ht]
    \centering
    \includegraphics[width=1\textwidth]{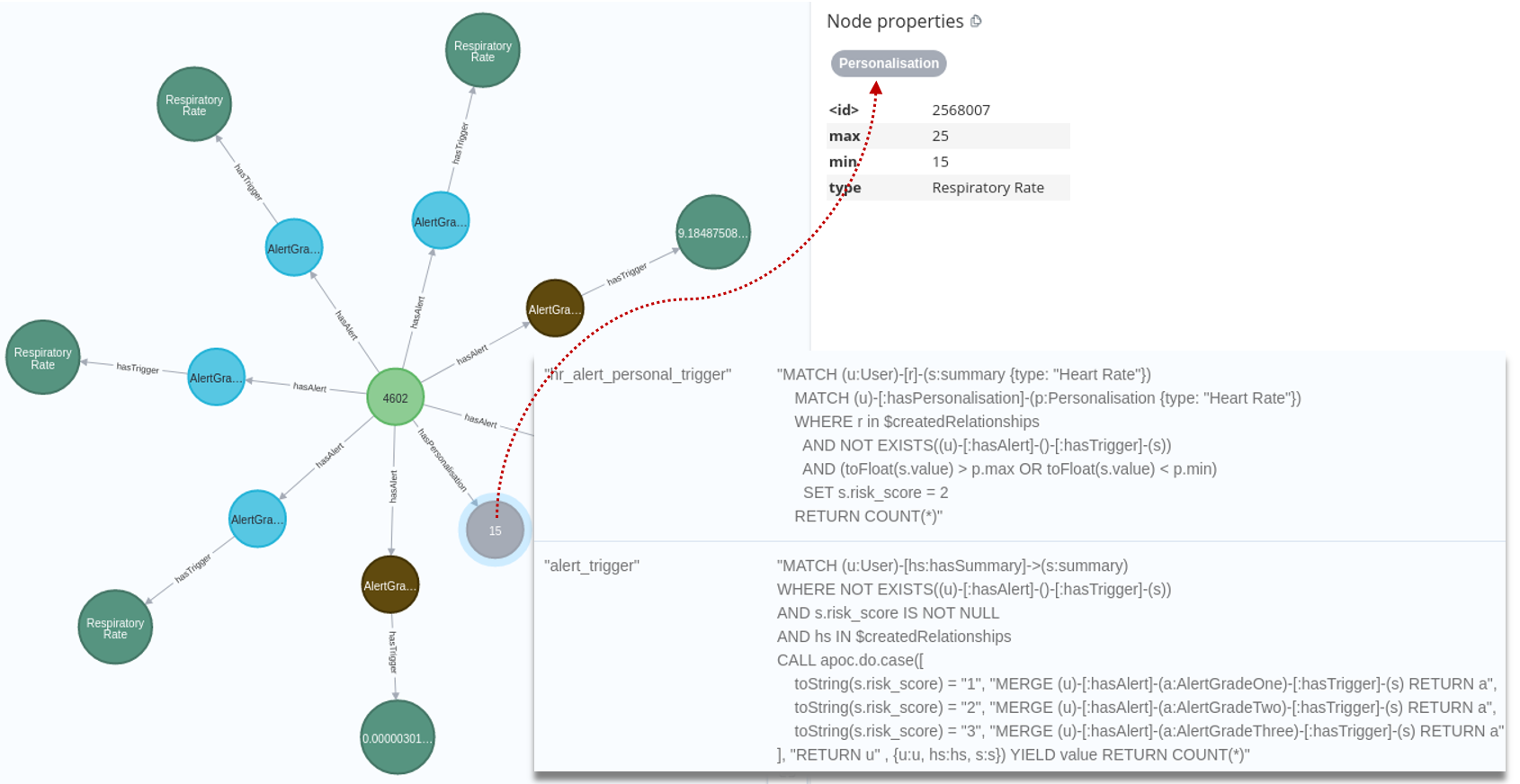}    
    \caption{Query on knowledge graph presenting measurements that triggered alerts, with the alert nodes and relationships also presented. A personalization relationship example is also featured, see the node property (personalization) indicated by the dashed arrow.}
    \label{fig:personalization}
\end{figure}

The system for COPD was shown to produce alerts based on the triggers that can be seen in Figure \ref{fig:personalization}. The query results in the assignment of risk scores based on the comparison between the value and the provided thresholds. These risk scores then trigger a final query that will create an alert node of the appropriate grade, a relationship between that alert and the user, and a relationship between the alert and the measurement node that triggered it. These alerts can act as actionable insights through the grading system of the alerts themselves, where each grade corresponds to a severity designed to be interpreted as a recommended response. A grade one alert would be a simple acknowledgement of a slightly abnormal reading, a grade two alert would be to suggest potential consultation or additional assessment, and a grade three alert is to indicate a need for immediate attention and urgent care for the patient.

\begin{figure}[ht]
    \centering
    \includegraphics[width=0.7\textwidth]{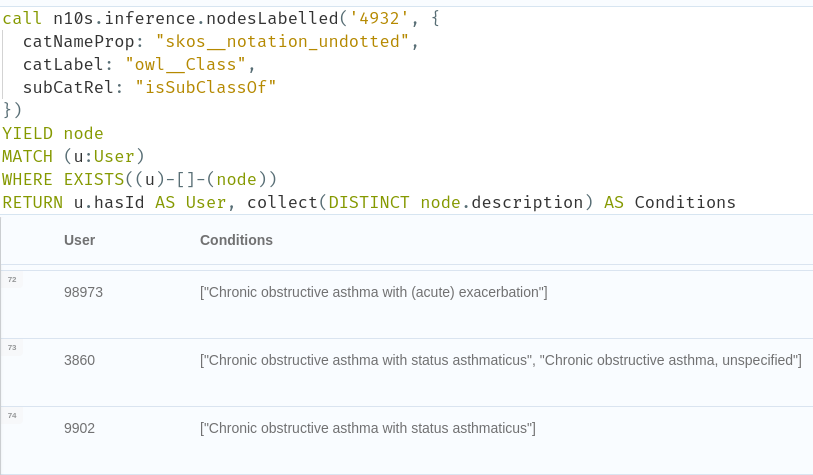}
    \caption{Querying PHKG on patient condition with subclass inferences. In the output below the query, we see user IDs and their diagnoses. While these three example patients have not been diagnosed with the exact condition queried, they exhibit conditions that are subclasses of it within the ontology. These conditions are of ICD9 codes 49322, 49321 and 49320 from the subclasses.}
    \label{fig:inference}
\end{figure}

Queries over PHKGs benefit greatly from the augmented domain knowledge stored within the graphs. As illustrated in Figure \ref{fig:inference}, we executed a query using ICD9 code ``4932'', which corresponds to ``Asthma with chronic obstructive pulmonary disease (COPD)''. Through the ``isSubClassOf'' relationship, the query infers the relevant conditions that the user may be interested in.

\section{\bf Conclusion}
\label{sec:CONCLUSIONS}

We have successfully applied our framework to construct personal health knowledge graphs (PHKGs) for monitoring COPD. By transforming personal health data and integrating it with relevant domain knowledge via ontology-based semantic mapping, our system achieves enhanced contextual understanding over the PHKGs, which can aid in automated alert generation, clinical decision support, and patient query. In future work, we will focus on improving the machine learning and multimodal analysis capabilities of the system, leveraging advanced algorithms such as graph neural networks, to enable more effective and accurate inferences.

\section*{\bf Acknowledgements}
\label{sec:Acknowledgements}
This work is partially funded by KESS II programme.
\section*{Availability of data and materials}
The code and demo of the PHKG for COPD is avialble at \href{https://github.com/Bluer01/COPH} {https://github.com/Bluer01/COPH}. 

\footnotesize
\bibliographystyle{IEEEtranN}
\bibliography{bibliography_CIBB_file.bib, biblio.bib} 
\normalsize

\end{document}